\documentclass[10pt,twocolumn,letterpaper]{article}

\usepackage{cvpr}
\usepackage{times}
\usepackage{epsfig}
\usepackage{graphicx}
\usepackage{amsmath}
\usepackage{amssymb}
\usepackage[font=small,labelfont=bf]{caption}
\usepackage{subcaption}
\usepackage{float}
\usepackage[hang,flushmargin]{footmisc}

\newcommand{\superscript}[1]{\ensuremath{^{\textrm{#1}}}}

\usepackage[pagebackref=true,breaklinks=true,letterpaper=true,colorlinks,bookmarks=false]{hyperref}

\newcommand\blfootnote[1]{%
  \begingroup
  \renewcommand\thefootnote{}\footnote{#1}%
  \addtocounter{footnote}{-1}%
  \endgroup
}

\cvprfinalcopy
\def\cvprPaperID{14}

\begin{document}

\title{Self-Supervised Feature Extraction for 3D Axon Segmentation}
\author{
    Tzofi Klinghoffer\superscript{1} \space\space\space\space
    Peter Morales\superscript{1} \space\space\space\space
    Young-Gyun Park\superscript{2} \space\space\space\space
    Nicholas Evans\superscript{2} \space\space\space\space \\
    Kwanghun Chung\superscript{2} \space\space\space\space
    Laura J. Brattain\superscript{1}
    \vspace{4pt} \\
    \superscript{1}MIT Lincoln Laboratory, Lexington, MA \\
    \superscript{2}MIT Institute for Medical Engineering and Science, Cambridge, MA \\
    {\tt\small \{Tzofi.Klinghoffer,Peter.Morales,BrattainL\}@ll.mit.edu
    \space\space
    \{ygpark,nbevans,khchung\}@mit.edu}
}
\maketitle

\blfootnote{DISTRIBUTION STATEMENT A. Approved for public release. Distribution is unlimited.
This material is based upon work supported by the National Institutes of Health (NIH). Any opinions, findings, conclusions or recommendations expressed in this material are those of the author(s) and do not necessarily reflect the views of the National Institutes of Health.
}

\begin{abstract}
Existing learning-based methods to automatically trace axons in 3D brain imagery often rely on manually annotated segmentation labels. Labeling is a labor-intensive process and is not scalable to whole-brain analysis, which is needed for improved understanding of brain function. We propose a self-supervised auxiliary task that utilizes the tube-like structure of axons to build a feature extractor from unlabeled data. The proposed auxiliary task constrains a 3D convolutional neural network (CNN) to predict the order of permuted slices in an input 3D volume. By solving this task, the 3D CNN is able to learn features without ground-truth labels that are useful for downstream segmentation with the 3D U-Net model. To the best of our knowledge, our model is the first to perform automated segmentation of axons imaged at subcellular resolution with the SHIELD technique. We demonstrate improved segmentation performance over the 3D U-Net model on both the SHIELD PVGPe dataset and the BigNeuron Project, single neuron Janelia dataset.

\end{abstract}
\section{Introduction}
Understanding brain connectivity is a long-standing goal in the neuroscience community. Recent advances in optical microscopy-based imaging methods, including CLARITY \cite{chung2013clarity}, Magnified Analysis of the Proteome (MAP) \cite{ku2016multiplexed}, and SHIELD \cite{park2019protection}, have enabled high-resolution, densely stained imaging of subcellular structures, such as axons. With these new imaging capabilities comes the need for new machine learning and image processing pipelines to detect axons and compute network connectivity. Existing methods, such as the work of Hernandez \etal\cite{hernandez2018learning}, first use a 3D CNN to segment the 3D volume, then skeletonize the detected axon voxels, and finally refine the axons with gap correction to mitigate segmentation errors and imaging artifacts. We focus our efforts on the first step of this pipeline, segmentation of 3D microscopy volumes.\par 
Machine learning methods for image and volume segmentation are often dependent on sufficient amounts of manually annotated training labels, which can be extremely difficult and time-consuming to acquire for 3D microscopy data. In this paper, we present a method to alleviate this dependency on labeled data through the use of self-supervised feature extraction.

\begin{figure}[!t]
\centering
\setlength\tabcolsep{2pt}
\begin{tabular}{cc}
\includegraphics[width = 1.5in]{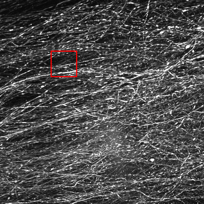} &
\includegraphics[width = 1.5in]{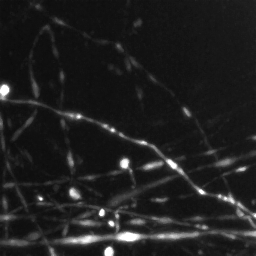} \\
\footnotesize{Full PVGPe Volume} & \footnotesize{Labeled PVGPe Subvolume}
\end{tabular}
\captionof{figure}{Max intensity projections of dense axons acquired using the SHIELD technique. Full volume is shown on the left ($2048 \times 2048 \times 1271$) with box indicating labeled subvolume ($256 \times 256 \times 206$), shown on the right. Due to axon density, labeling is time-consuming and difficult, motivating our work to utilize unlabeled data during training.}
\label{figure:mip}
\end{figure}

\subsection{Contributions}
We present a self-supervised approach to utilize unlabeled 3D microscopy data for axon segmentation. The encoder of the 3D U-Net model proposed by Çiçek \etal\cite{cciccek20163d} is first pre-trained on an auxiliary task using unlabeled data and then the entire 3D U-Net model is fine-tuned on the axon segmentation task using labeled data. Using this framework, we present the proposed auxiliary task, which involves reordering slices in each training subvolume so that the tube-like structure of axons is corrupted. The auxiliary task is to then use the 3D U-Net encoder and an auxiliary classifier to predict the permutation that was used to reorder the slices of each input subvolume, encouraging the encoder to learn features related to axon structure. We also discuss a component of our self-supervised learning loss function, dubbed information weighting. Information weighting is used to prevent penalization of poor performance on training samples with few or no axons. Finally, we demonstrate the benefits of our methods on two optical microscopy datasets, one containing dense axons from the mouse hippocampus imaged with SHIELD and one containing single neurons from the adult Drosophila nervous system.\par
Specifically, we make the following contributions:
\begin{itemize}
  \setlength{\itemsep}{0mm}
  \item A self-supervised auxiliary task for 3D U-Net architectures that improves axon segmentation performance over existing methods.
  \item A methodology for improving training performance on datasets with variable axon densities, which we refer to as information weighting.
  \item An empirical evaluation of the impact of the proposed auxiliary task on axon segmentation performance.
\end{itemize}

\section{Related Work}

\subsection{Brain Mapping }
Mapping of the brain has advanced rapidly in recent years with the release of the BigNeuron Project, led by the Allen Institute, which includes neuronal reconstruction algorithms and publicly available, single neuron datasets to benchmark against \cite{peng2015bigneuron}. Despite these advances, tracing of both single neurons and dense axons remains a challenging task. Existing methods can be generally classified as either image processing-based or deep learning-based.\par
Image processing-based methods often start with generating an over-segmentation with watershed \cite{lee2015recursive,vincent1991watersheds}, followed by region proposal merging \cite{funke2017deep,lee2017superhuman,nunez2013machine}. Common methods for region proposal merging include simulated annealing \cite{ren2003learning} and hierarchical clustering \cite{jain2011learning}. Deep learning-based methods for brain mapping often use segmentation techniques to distinguish axons and neurons from other anatomy \cite{li20193d,li2017deep,wang2019multiscale}. U-shaped architectures, such as 3D U-Net \cite{cciccek20163d} are most popular for this task. Various improvements have been suggested for the use of 3D U-Net to trace axons and neurons \cite{wang2019multiscale}. Other deep-learning-based methods use flood-filling to trace neurons outward from an initial neuron voxel \cite{januszewski2018high}. While many of these methods operate on electron microscopy data, we focus on automated segmentation of axons present in optical microscopy data. 

\begin{figure}[t] 
\begin{center}
   \includegraphics[width=1\linewidth]{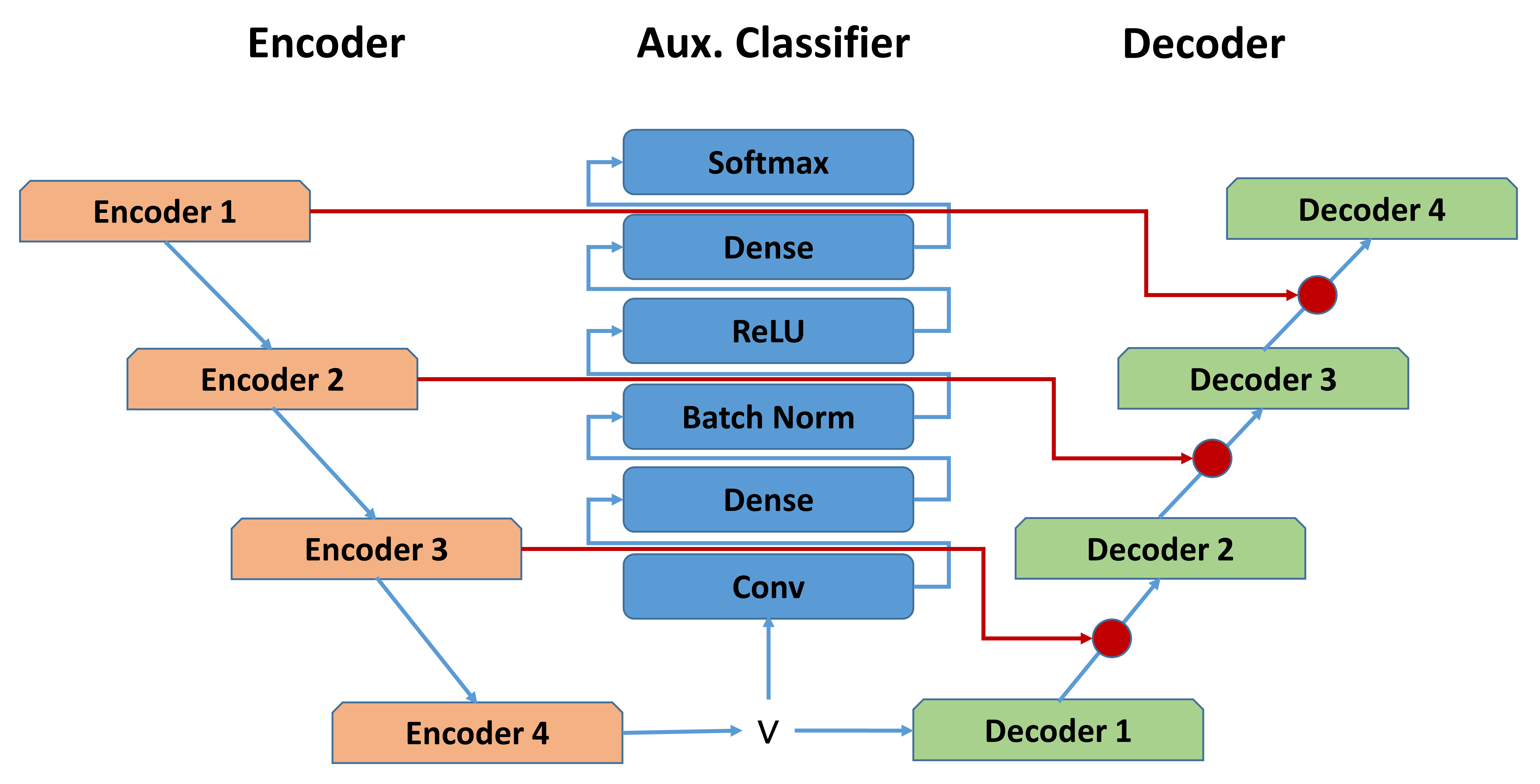}
\end{center}
   \captionof{figure}{Our proposed model. The encoder and auxiliary classifier are first trained to learn the proposed self-supervised auxiliary task. Then, 3D U-Net, composed of the pre-trained encoder and randomly initialized decoder, is trained on the segmentation task using the limited labels available.}
\label{figure:model}
\end{figure}

\subsection{Self-Supervised Learning}
Self-supervised learning (SSL) is a form of unsupervised learning that has become widely popular in recent years \cite{doersch2015unsupervised,DBLP:journals/corr/abs-1708-07860,DBLP:journals/corr/abs-1803-07728,misra2016shuffle,norooziECCV16}. SSL methods leverage large unlabeled datasets by withholding portions of the dataset and creating auxiliary tasks whose goal is to predict the withheld portion. Once trained on the auxiliary task, relevant learned representations can be transferred to a target supervised learning task. These approaches are most beneficial in data regimes where labeled data is sparse \cite{su2019does,zhai2019largescale}, which is often the case for medical datasets. Additionally, these approaches have the benefit of  learning features from the auxiliary task that can lead to more robust models \cite{DBLP:journals/corr/abs-1906-12340}. This robustness is likely due to the auxiliary task forcing new semantic features to be learned that are applicable to the target task \cite{raghu2019rapid}.\par

Several SSL methods have been proposed for classification, regression, and segmentation of 3D medical volumes. Often these methods extend auxiliary tasks proposed for 2D images, such as the work of Zhuang \etal \cite{zhuang2019self}, which extended the jigsaw puzzle task by Noroozi and Favaro \cite{norooziECCV16} into the task of solving a Rubik's cube. Instead of reordering patches in a 2D image, the volume is broken into cubes that are scrambled and rotated, requiring the CNN to predict both the cube order and rotation. Features learned in the Rubik's cube task were shown to be helpful for both classification and segmentation of CT data. Zhang \etal \cite{zhang2017self} proposed a slice ordering task for body part recognition in CT data, but did not use a 3D CNN and instead built a 2D CNN classifier to identify the relative position of pairs of input slices randomly sampled from the input 3D volume. They observed that this task forces the CNN to learn spatial context information helpful for the downstream, body recognition task. Inspired by these approaches, we propose an auxiliary task to improve axon segmentation. Because our experiments indicated that learning the Rubik's cube task for SHIELD data is highly difficult, we focused our efforts on re-formulating the slice ordering task to maximize axon segmentation performance.\par

\section{Proposed Method}
In our work, we propose incorporating self-supervised learning into the popular 3D U-Net segmentation model. First, the 3D U-Net encoder is pre-trained on an auxiliary task that requires no labels. Once trained, the entire 3D U-Net model, including the pre-trained encoder and the randomly initialized decoder, can be trained on the axon segmentation task using limited labeled data. Figure \ref{figure:model} illustrates our model, including the auxiliary classifier used to solve the auxiliary task.

\begin{figure*}
\centering
    \setlength\tabcolsep{2pt}
	\begin{tabular}{ccccc}
    	\includegraphics[width = 1.32in, keepaspectratio]{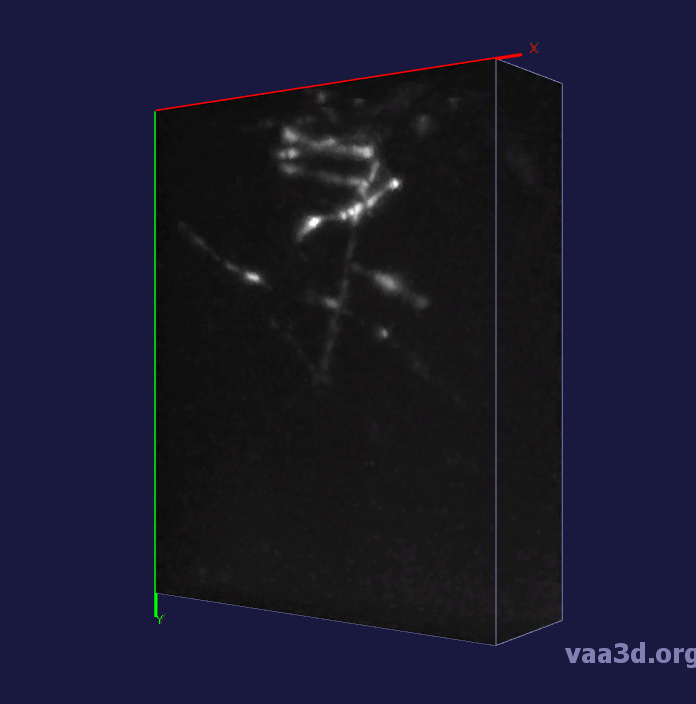} &
    	\includegraphics[width = 1.32in, keepaspectratio]{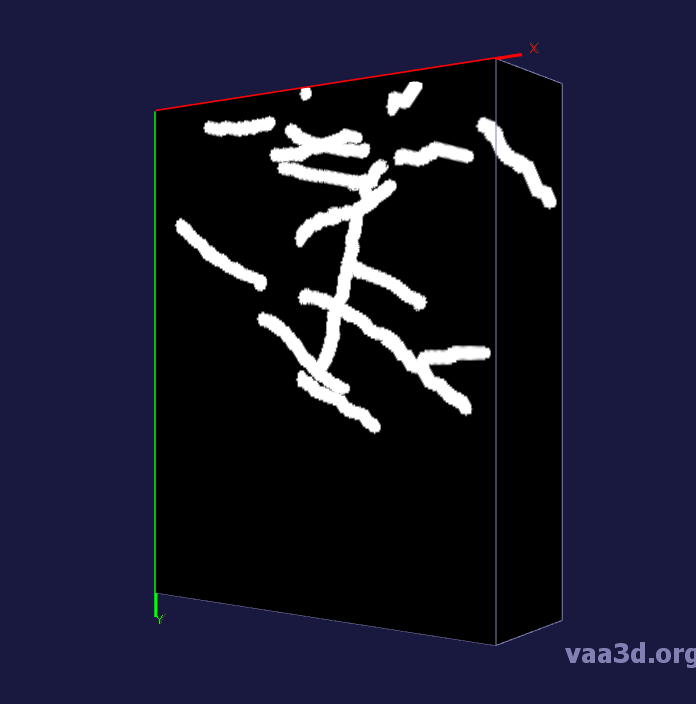} &
    	\includegraphics[width = 1.32in, keepaspectratio]{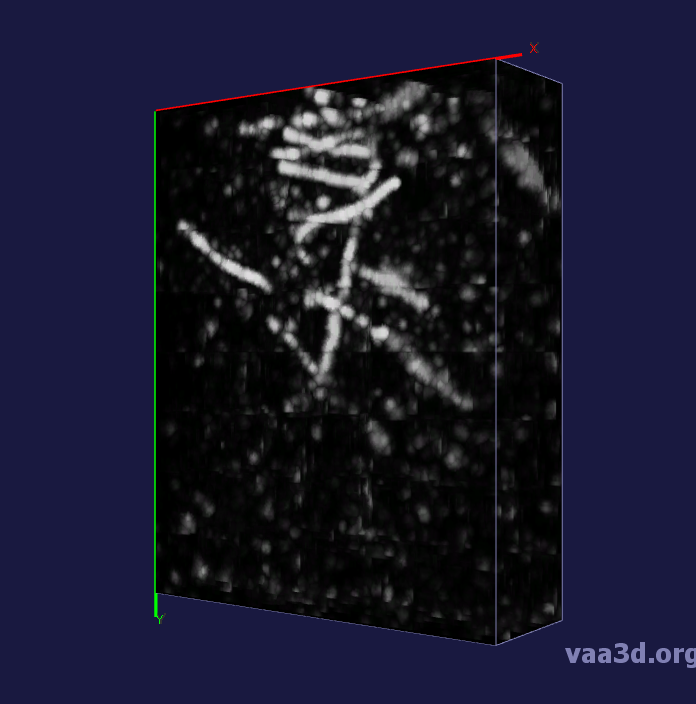} &
    	\includegraphics[width = 1.32in, keepaspectratio]{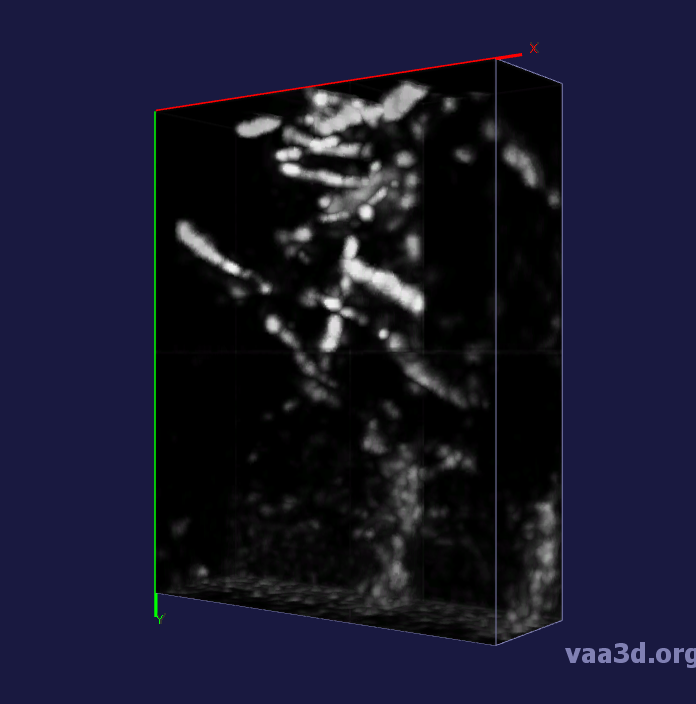} &
    	\includegraphics[width = 1.32in, keepaspectratio]{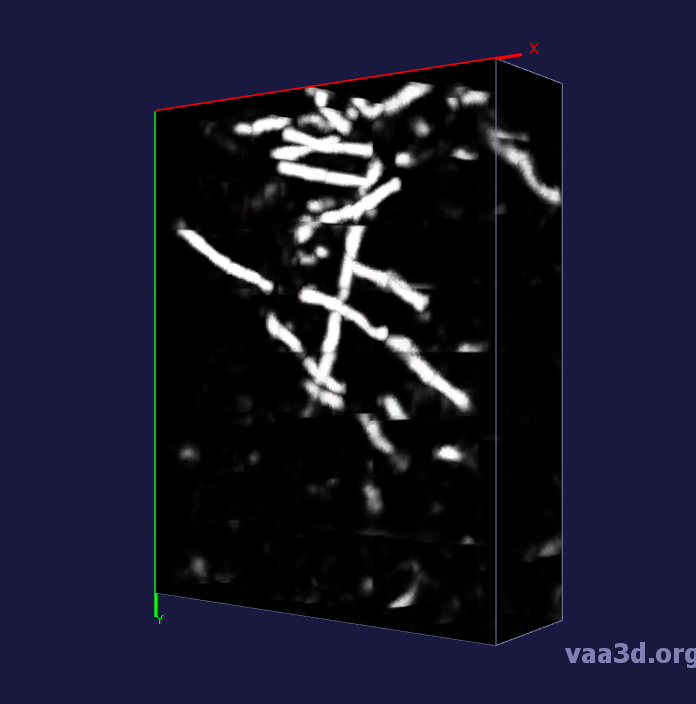} \\
    	\includegraphics[width = 1.32in, keepaspectratio]{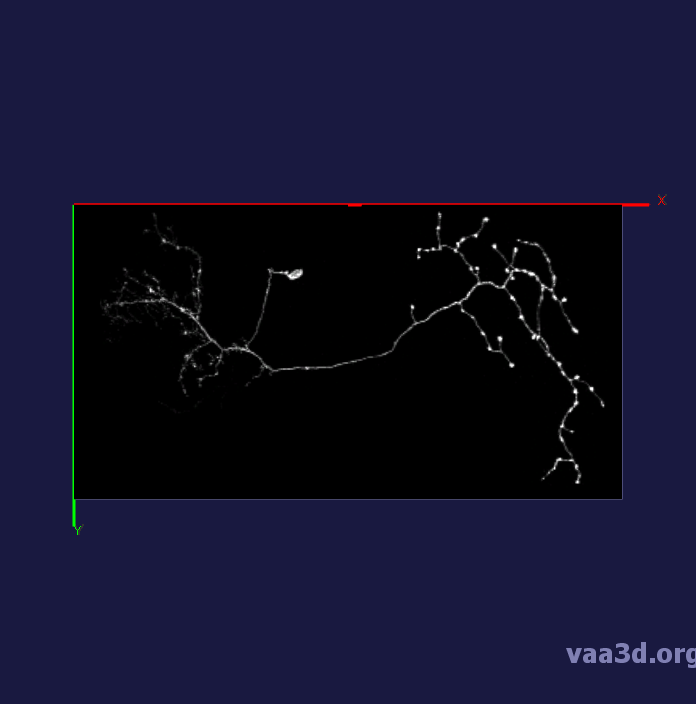} &
    	\includegraphics[width = 1.32in, keepaspectratio]{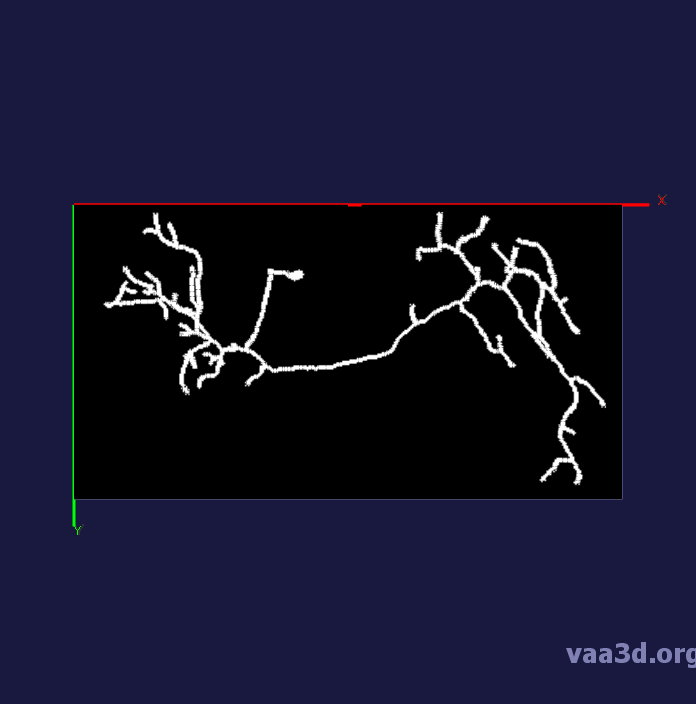} & 
    	\includegraphics[width = 1.32in,keepaspectratio]{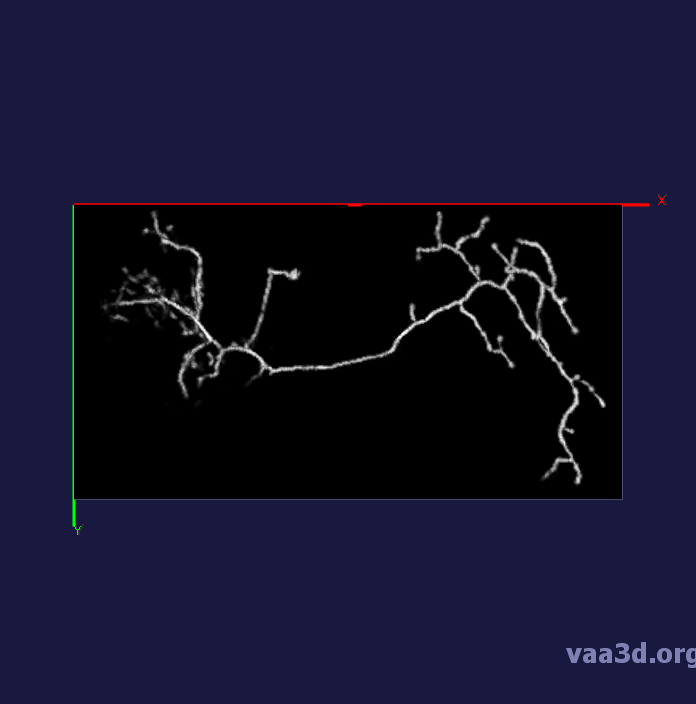} &
    	\includegraphics[width = 1.32in, keepaspectratio]{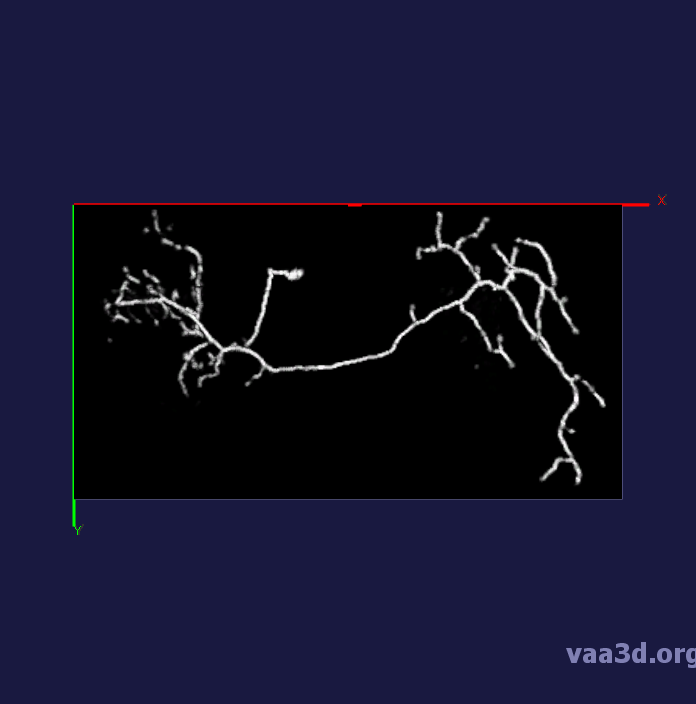} &
    	\includegraphics[width = 1.32in, keepaspectratio]{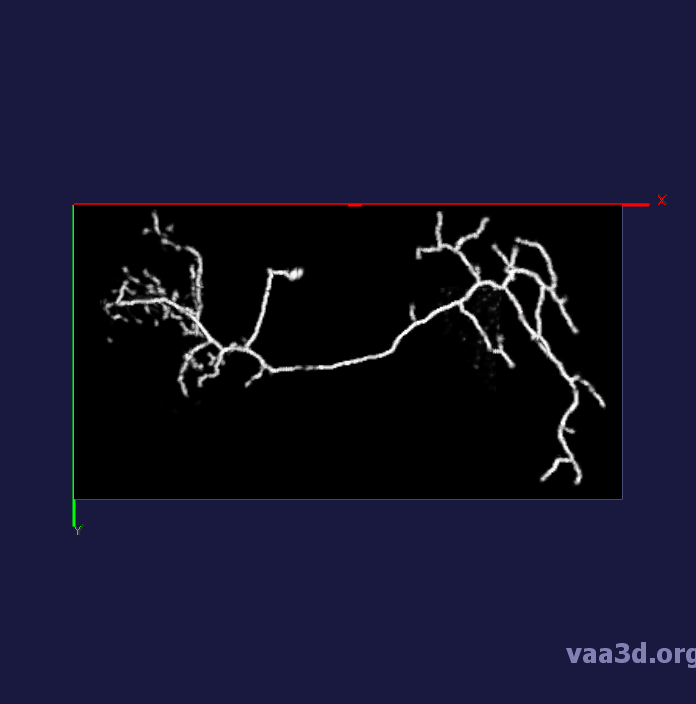} \\
      	Input & Truth & 2D U-Net & 3D U-Net & Proposed
    \end{tabular}
    \captionof{figure}{Examples of segmentation masks generated by 2D U-Net, 3D U-Net, and the proposed approach. On the top are results from the SHIELD PVGPe dataset and on the bottom are results from the Janelia dataset. Results are not thresholded and pixel intensities embed model confidence. The proposed approach generates higher confidence and more precise segmentation predictions.}
\label{figure:segmentation}
\end{figure*}

\subsection{Self-Supervised Auxiliary Task}
Our proposed self-supervised auxiliary task is motivated by the tube-like structure of axons. We treat the slices of each 3D volume as frames in a video that can be reordered, as is done in the shuffle and learn task proposed by Misra \etal\cite{misra2016shuffle}. By reordering the slices, the structure of axons in the volume is corrupted. Rather than tasking the CNN with predicting whether the slices are in the correct order, we instead task it with classifying the permutation used to reorder the slices. We opt for this more challenging task based on previous work that has shown that more difficult auxiliary tasks produce more useful feature representations \cite{noroozi2018boosting,su2019does}.\par
The proposed auxiliary task requires that each training sample be split into slices along an axis. We chose to use the \textit{z}-axis because this is the primary direction in which axons and neurons move in the datasets used. If each training sample has a size $\mathrm{X}\times\mathrm{Y}\times\mathrm{Z}$, then there are $\mathrm{Z}!$ possible permutations of slices along the \textit{z}-axis. In our experiments, we chose $\mathrm{Z} = 8$. We first construct a set of $N$ random permutations of the $\mathrm{Z}$ slices and enforce a minimum hamming distance between each selected permutation as was done in the work of Noroozi and Favaro \cite{norooziECCV16}. Enforcing a minimum distance between permutations reduces ambiguity in the auxiliary task by guaranteeing no permutations are highly similar. During training, a random permutation from the constructed set is applied to each training sample. The CNN is then tasked with predicting the index, 0 to $N$, of the permutation used. For example, the original order of slices in a training sample can be represented as $[0,1,2,3,4,5,6,7]$ and a possible permutation is $[5,2,1,7,0,4,6,3]$. We used a one-hot encoding to label each training sample where $argmax$ of the encoding is the index to the permutation applied to the sample.\par
We trained the encoder of the 3D U-Net model and an auxiliary classifier together to learn the proposed auxiliary task. We chose to use the 3D U-Net encoder to enable downstream transfer of the learned features into the full 3D U-Net model used for segmentation. Whereas the output of the encoder would normally be input to the decoder, during training of the auxiliary task, we instead pass it into the auxiliary classifier, composed of two fully connected layers and a final softmax activation layer. The output of the auxiliary classifier is a one-hot encoding of length $N$. The auxiliary classifier architecture is shown in Figure \ref{figure:model}.

\subsection{Self-Supervised Loss}
The proposed auxiliary task is trained using cross-entropy loss and information weighting. We define information weighting as the ratio of the sum of the training sample to the sum of the whole volume that the sample was drawn from. This can be stated as
\begin{equation} \label{loss}
\mathcal{L}(y, \hat{y}, x_{j}, x) = \frac{\sum_{i=1}^{V_j} x_{j}}{\sum_{i=1}^V x} \times \left( -\sum_{i=1}^N y_{i} \times \log\hat{y_{i}} \right)
\end{equation}
In Equation \ref{loss}, $y$ is the index of the permutation, $\hat{y}$ is the predicted index of the permutation, $x_{j}$ is the flattened, permuted subvolume, and $x$ is the flattened full volume from which $x_{j}$ was drawn. The weighting of the cross-entropy is motivated by the observation that axons have a higher voxel intensity than the background. Therefore, we expect subvolumes with more axons and hence more information to have a higher sum of voxel intensities than subvolumes with fewer axons. Because only axons are visible in microscopy data generated with CLARITY, MAP, and SHIELD, if there are no axons present in a selected subvolume, then there is no information available to predict slice order or structure. Thus, the task of permutation classification becomes impossible. Information weighting is introduced to avoid penalizing poor performance in this situation. We assume uniform noise across whole volumes.
 
\subsection{Implementation \& Training Details}
Our implementation was developed in PyTorch and is available online\footnote{\url{https://github.com/tzofi/ssl-for-axons}}. We used ADAM \cite{kingma2014adam} as an optimizer and an initial learning rate of $1\times 10^{-3}$ for both the auxiliary and segmentation tasks. Early stopping was used to prevent overfitting and training ended after 100 epochs without improvement in validation loss. During training of both the auxiliary and segmentation tasks, subvolumes of a specified size were randomly sampled from the training volumes. During validation and inference, a sliding window of the same size was used to sample across the volume, ensuring the same subvolumes were sampled each time validation or testing was done. We utilized the 3D U-Net implementation by Wolny \etal\cite{Wolny2020.01.17.910562}. For the auxiliary task, the number of slices per sample was set to eight in each experiment. For the segmentation task, binary cross-entropy loss was used and rotation was randomly applied to each axis of training samples for augmentation. 

\section{Experimental Results}
\subsection{Datasets}
\subsubsection{SHIELD PVGPe Dataset}
We trained and evaluated our models on a microscopy dataset called PVGPe imaged from a one mm-thick mouse brain tissue that includes areas of globus pallidus externa, globus pallidus interna, substantia nigra reticulata, and subthalamic nucleus mouse. The tissue was stained with calretinin antibody immunostaining, prepared using the SHIELD technique \cite{park2019protection}, and imaged using a light-sheet imager (SMARTSPIM, Lifecanvas) after 3X tissue expansion. The acquired PVGPe image volume is $2048 \times 2048 \times 1271$ voxels, with a voxel resolution of $0.65 \times 0.65 \times 2\ \mu m$ in \textit{x, y, z}. The labeled subvolume is $256 \times 256 \times 206$. The max intensity projections for both are shown in Figure \ref{figure:mip}. We first preprocessed the PVGPe volumes by clipping the lowest and highest values, applying a median filter, and scaling values between zero and one with min-max normalization. We then split the labeled PVGPe subvolume as follows for segmentation training: $128 \times 256 \times 206$ for training, $64 \times 256 \times 206$ for validation, and $64 \times 256 \times 206$ for testing. When pre-training the 3D U-Net encoder on the auxiliary task,  the same training subvolume was used. We also experimented with increasing the amount of training data for the auxiliary task by including an additional, equally sized, training subvolume. These experiments are called Proposed and Proposed+, respectively. The additional subvolume used was sampled from an area adjacent to the original training subvolume to reduce variability in axon density that is present throughout the volume.\par

\subsubsection{Janelia Dataset}
We also evaluated the proposed approach on the Janelia dataset from the BigNeuron Project \cite{peng2015bigneuron}, consisting of optical microscopy data of single neurons from the adult Drosophila nervous system. This dataset includes 42 volumes. 35 volumes were used for self-supervised training and 18 of those 35 were also used to train the supervised segmentation model. Of the remaining seven volumes, three were used for validation and four were used for testing. We scaled all volumes between zero and one using min-max normalization. As in Wang \etal\cite{wang2019multiscale}'s prior work utilizing this dataset, we used a sample size of $128 \times 128 \times 64$ for training and inference.

\begin{table*}[h]
\centering
\begin{tabular}{p{0.24\linewidth}p{0.15\linewidth}p{0.11\linewidth}p{0.11\linewidth}p{0.11\linewidth}p{0.11\linewidth}}
\hline
Method & Sample Size & AUC Mean & AUC Std. & F1 Mean & F1 Std. \\
\hline
\hline
3D CNN (Hernandez \cite{hernandez2018learning}) & $19 \times 19 \times 19$ & 0.4305 & -- & 0.4710 & -- \\
\hline
2D U-Net (Ronneberger \cite{ronneberger2015u}) & $32 \times 32 \times 1$ &  \textbf{0.4942} &  \textbf{0.0125} & 0.5179 & 0.0106 \\
3D U-Net (Çiçek \cite{cciccek20163d}) & $32 \times 32 \times 32$ & 0.4239 & 0.0374 & 0.4913 & 0.0453 \\
Proposed Approach & $32 \times 32 \times 32$ & 0.4652 & 0.0294 & 0.5845 & \textbf{0.0075} \\
Proposed+ Approach & $32 \times 32 \times 32$ & 0.4661 & 0.0294 & \textbf{0.5884} & 0.0097 \\
\hline
2D U-Net (Ronneberger \cite{ronneberger2015u}) & $128 \times 128 \times 1$ & 0.2756 & 0.0344 & 0.3800 & 0.0485 \\
3D U-Net (Çiçek \cite{cciccek20163d}) & $128 \times 128 \times 64$ & 0.3397 & \textbf{0.0322} & 0.5236 & 0.0636 \\
Proposed Approach & $128 \times 128 \times 64$ & 0.4088 & 0.0601 & 0.5576 & 0.0251 \\
Proposed+ Approach & $128 \times 128 \times 64$ & \textbf{0.4523} & 0.0626 & \textbf{0.5718} & \textbf{0.0150} \\
\hline
\end{tabular}
\captionof{table}{Average performance of models over six trials on the SHIELD PVGPe dataset containing dense axons. The proposed approach consists of pre-training the 3D U-Net encoder on an auxiliary task using unlabeled data and then training the entire 3D U-Net on the segmentation task. Proposed+ utilizes twice as much training data for the auxiliary task and achieves the highest AUC among 3D models and highest F1 score among all models. F1 standard deviation (std) is also lower among models trained with the proposed approach.}
\label{table:segresults}
\end{table*}

\begin{table*}[h]
\centering
\begin{tabular}{p{0.24\linewidth}p{0.15\linewidth}p{0.11\linewidth}p{0.11\linewidth}p{0.11\linewidth}p{0.11\linewidth}}
\hline
Method & Sample Size & AUC Mean & AUC Std. & F1 Mean & F1 Std. \\
\hline
\hline
2D U-Net (Ronneberger \cite{ronneberger2015u}) & $128 \times 128 \times 1$ &  0.6782 &  \textbf{0.0131} & 0.6603 & \textbf{0.0050} \\
3D U-Net (Çiçek \cite{cciccek20163d}) & $128 \times 128 \times 64$ & 0.6796 & 0.0164 & 0.6965 & 0.0069 \\
Proposed Approach & $128 \times 128 \times 64$ & \textbf{0.6971} & 0.0215 & \textbf{0.7062} & 0.0053 \\
\hline
\end{tabular}
\captionof{table}{Performance of models on the Janelia dataset containing single neurons. Each model was trained and evaluated six times to compute mean and standard deviation (std) values.}
\label{table:janelia-results}
\end{table*}

\subsection{Segmentation Experiments}
In this section, we compare the performance of the proposed self-supervised 3D U-Net model with other methods, including past work by Hernandez \etal\cite{hernandez2018learning}, the 2D U-Net model developed by Ronneberger \etal\cite{ronneberger2015u}, and the 3D U-Net model developed by Çiçek \etal\cite{cciccek20163d}. We used Area Under the Curve (AUC) and top F1 score as metrics to measure voxel-level segmentation accuracy. AUC was computed with threshold increments of 0.05. For each experiment, we trained the model six times and report both the mean and standard deviation of the AUC and F1 metrics.\par

For all experiments that included pre-training the 3D U-Net encoder for the auxiliary task, we used eight slices along the \textit{z}-axis per sample and generated ten permutations, tasking the auxiliary classifier with predicting which of the ten permutations was used to reorder each sample.

\subsubsection{SHIELD PVGPe Dataset Results}
Results are shown in Table \ref{table:segresults} and indicate that the proposed self-supervised method outperforms both the 3D CNN used by Hernandez \etal\cite{hernandez2018learning} and the 2D and 3D U-Net models. We carried out two sets of segmentation experiments, one with $32 \times 32 \times 32$ sized samples and one with $128 \times 128 \times 64$ sized samples. During these experiments, the 2D U-Net model was trained with one slice per sample, i.e. $32 \times 32 \times 1$ and $128 \times 128 \times 1$, and the self-supervised auxiliary task used in the proposed method was trained with eight slices per sample, i.e. $32 \times 32 \times 8$ and $128 \times 128 \times 8$. Self-supervised pre-training resulted in improved performance in both sets of experiments and had the most significant impact when larger, and thus fewer, samples were used. For larger sample sizes, the self-supervised approach improved AUC by over 10\% and F1 by nearly 5\% over the 3D U-Net model. Despite the proposed approach also achieving the highest F1 score when smaller sample sizes were used, the 2D U-Net achieved the highest AUC. We argue F1 as the more important metric for axon segmentation as it indicates the ability of the model to maximize recall and precision simultaneously, and has a larger impact on qualitative results, as shown in Figure \ref{figure:segmentation}. In each set of experiments, we trained the 2D and 3D U-Net models, and two variants of the proposed approach, Proposed and Proposed+. Proposed used only the segmentation training subvolume to train the auxiliary task, whereas Proposed+ included an additional training subvolume to train the auxiliary task. We observed that even if all data are labeled and can be used to train the segmentation model, there is still benefit in first learning the auxiliary task. Furthermore, increasing the amount of unlabeled training data for the auxiliary task leads to increased downstream segmentation accuracy, as Proposed+ achieved the highest F1 score in both sets of experiments.\par

All experiments that included self-supervised pre-training of the encoder had a lower standard deviation in top F1 score than 3D U-Net, indicating that random weight initialization had less of an impact on performance when the encoder was pre-trained on the auxiliary task. We also observed that models that were first pre-trained generated predictions that were less noisy and higher confidence than all other models evaluated, as shown in Figure \ref{figure:segmentation}.\par

In evaluating the accuracy of the proposed model on the auxiliary task, we conducted 50 trials to account for randomness in permutation selection per input subvolume. The resulting mean test accuracy was 67\%, indicating that the network was able to learn the auxiliary task effectively.

\subsubsection{Janelia Dataset Results}
When segmenting single neurons in the Janelia dataset, we used sample sizes of $128 \times 128 \times 8$ for self-supervised pre-training and $128 \times 128 \times 64$ for segmentation. The 2D U-Net model was trained with sample sizes of $128 \times 128 \times 1$ in these experiments. The proposed self-supervised approach resulted in incremental improvements in both AUC and F1 score over both the 2D and 3D U-Net models during six trials, as shown in Table \ref{table:janelia-results}. The mean AUC improved approximately 1\% and the mean F1 score improved 1.75\% over the 3D U-Net model. While still beneficial to performance on the Janelia dataset, the proposed approach appears to have a larger impact on more challenging segmentation tasks that have less labeled data, as was the case with the SHIELD PVGPe dataset. Resulting segmentation masks for the Janelia dataset are shown in Figure \ref{figure:segmentation}.

\section{Discussion}
Our results indicate that features learned from a self-supervised auxiliary task can be used to improve axon segmentation. In analyzing the precision and recall values that contribute to the AUC for each model, we observed that, in comparison to both the 2D and 3D U-Net models, the proposed approach consistently yields higher precision. This finding suggests that incorporating the proposed auxiliary task may help reduce false positives and noise, which is further supported by the segmentation masks, shown in Figure \ref{figure:segmentation}. We also noticed that there is a notable amount of variability in training these models, especially due to the small amount of labeled training data. We capture this variability by training each model six times and reporting the standard deviation in test performance. Additional trials could lead to incremental changes in results. Furthermore, the incorporation of additional labeled data for training, validation, and testing would decrease variability and better capture the overall benefit of the proposed models.

\section{Conclusion}
This paper proposed the use of self-supervised learning to extract features from unlabeled microscopy data that can be utilized for improved axon segmentation. Our work is the first to demonstrate automated segmentation of axons in data imaged with the SHIELD technique. We focused on an auxiliary task centered around reordering slices in each input 3D subvolume and constraining the 3D U-Net encoder and an auxiliary classifier to predict the permutation used for the reordering. By learning to solve this task, the 3D U-Net encoder learns high-level features regarding axon structure that can be transferred into the full 3D U-Net model for segmentation. We demonstrated that this approach results in higher segmentation accuracy than achieved with existing models on both the dense axon SHIELD PVGPe dataset and the single neuron Janelia dataset from the BigNeuron Project. The proposed approach can easily be incorporated into the widely used 3D U-Net to improve axon segmentation in optical microscopy data.

{\small
\bibliographystyle{ieee_fullname}
\bibliography{neuron_paper}
}

\end{document}